# A New Fuzzy Approach for Dynamic Load Balancing Algorithm

Abbas Karimi[1,2,3], Faraneh Zarafshan [1,3], Adznan b. Jantan[1],
A.R. Ramli[1], M. Iqbal b.Saripan[1]

[1]Department of Computer Systems Engineering, Faculty of Engineering, UPM, Malaysia
[2]Computer Department, Faculty of Engineering, IAU, Arak, Iran
[3]Young Researchers' Club, IAU, Arak, Iran
akarimi @iau-arak.ac.ir

*Abstract*— Load balancing is the process of improving the Performance of a parallel and distributed system through is distribution of load among the processors[1-2]. Most of the previous work in load balancing and distributed decision making in general, do not effectively take into account the uncertainty and inconsistency in state information but in fuzzy logic, we have advantage of using crisps inputs. In this paper, we present a new approach for implementing dynamic load balancing algorithm with fuzzy logic, which can face to uncertainty and inconsistency of previous algorithms, further more our algorithm shows better response time than round robin and randomize algorithm respectively 30.84% and 45.45%.

## I. INTRODUCTION

Distributed computing systems have become a natural setting in many environments for business and academia. This is due to the rapid increase in processor and/or memory hungry applications coupled with the advent of low-cost powerful workstations[3].

In a typical distributed system setting, tasks arrive at the different nodes in a random fashion. This causes a situation of non-uniform loading across the system nodes to occur. Loading imbalance is observed by the existence of nodes that are highly loaded while others are lightly loaded or even idle. Such situations are harmful to the system performance in terms of the mean response time of tasks and resource utilization[3].

A system [4-5] of distributed computers with tens or hundreds of computers connected by high-speed networks has many advantages over a system that has the same standalone computers. A distributed system provide the resource sharing as one of its major advantages, which provide the better performance and reliability than any other traditional system in the same conditions[1].

Section II describes the load balancing and the kinds of its models. In section III, we explain and demonstrate our model, then in section IV, we explain the methodology and fuzzy rules. The evaluation of performance is inspected in section V and finally we describe the conclusion.

## II. LOAD BALANCING

In computer networking, load balancing is a technique to spread work between two or more computers, network links, CPUs, hard drives, or other resources, in order to get optimal resource utilization, throughput, or response time. Using multiple components with load balancing, instead of a single component, may increase reliability through redundancy. Load balancing attempts to maximize system throughput by keeping all processors busy Load balancing is done by migrating tasks from the overloaded nodes to other lightly loaded nodes to improve the overall system performance.

Load balancing algorithms are typically based on a *load index,* which provides a measure of the workload at a node relative to some global average, and four *policies,* which govern the actions taken once a load imbalance is detected[6]. The load index is used to detect a load imbalance state. Qualitatively, a load imbalance occurs when the load index at one node is much higher (or lower) than the load index on the other nodes. The length of the CPU queue has been shown to provide a good load index on timeshared workstations when the performance measure of interest is the average response time[7-8]. In the case of multiple resources (disk, memory, etc.), a linear combination of the length of all the resource queues provided an improved measure, as job execution time may be driven by more than CPU cycles[9-10] .
The four policies that govern the action of a load-balancing algorithm when a load imbalance is detected deal with information, transfer, location, and selection. The *information*
Policy is responsible for keeping up-to-date load information about each node in the system. A global information policy provides access to the load index of every node, at the cost of additional communication for maintaining accurate information[5, 10].
The *transfer* policy deals with the dynamic aspects of a system. It uses the nodes' load information to decide when a node becomes eligible to act as a sender (transfer a job to another node) or as a receiver (retrieve a job from another node). Transfer policies are typically threshold based. Thus, if the load at a node increases beyond a threshold ., the node becomes an eligible sender. Likewise, if the load at a node drops below a threshold, the node becomes an eligible receiver

The *location* policy selects a partner node for a job transfer transaction. If the node is an eligible sender, the location policy seeks out a receiver node to receive the job selected by the selection policy (described below). If the node is an eligible receiver, the location policy looks for an eligible sender node[10].

Once a node becomes an eligible sender, a *selection* policy is used to pick which of the queued jobs is to be transferred to the receiver node. The selection policy uses several criteria to evaluate the queued jobs. Its goal is to select a job that reduces the local load, incurs as little cost as possible in the transfer, and has good affinity to the node to which it is transferred. A common selection policy is *latest-job arrived* which selects the job which is currently in last place in the work queue[10].

There are two types of load balancing algorithms:

### A. Static Load-Balancing

In this method, the performance of the nodes is determined at the beginning of execution. Then depending upon their performance the workload is distributed in the start by the master node. The slave processors calculate their allocated work and submit their result to the master. A task is always executed on the node to which it is assigned that is static load balancing methods are non-preemptive. A general disadvantage of all static schemes is that the final selection of a host for process allocation is made when the process is created and cannot be changed during process execution to make changes in the system load[1]. Major load balancing algorithms are Round Robin[11] and Randomized Algorithms[12], Central Manager [13]Algorithm and Threshold[1, 14] Algorithm.

### B. Dynamic Load-Balancing

It differs from static algorithms in that the workload is distributed among the nodes at runtime. The master assigns new processes to the slaves based on the new information collected[4, 15]. Unlike static algorithms, dynamic algorithms allocate processes dynamically when one of the processors becomes under loaded. Instead, they are buffered in the queue on the main host and allocated dynamically upon requests from remote hosts[1]. This method is consisted of Central Queue Algorithm and Local Queue Algorithm[16].

Load balancing algorithms work on the principle that in which situation workload is assigned, during compile time or at runtime. Comparison shows that static load balancing algorithms are more stable compare to dynamic. It is also ease to predict the behavior of static, but at the same time, dynamic distributed algorithms are always considered better than static algorithms[1].

### III. SYSTEM MODEL

We have a distributed network consists of n node which every node may be a complex combination of multiple types of resources (CPUS, memory, disks, switches, and so on) and the physical configurations of resources for each node may be heterogeneous. This heterogeneity can be manifested in two ways[17]. The amount of a given resource at one node site may be quite different from the configuration of a node at another site. Additionally, nodes may have different balance of each resource. For example, one node may have a (relatively) large memory with respect to its number of CPUs while another node may have a large number of CPUs with less memory [18-19].

As in Fig. 1 illustrated, our system model is involved Routing table, Load index, Cost table and a fuzzy controller, which manages Load balancing of system.

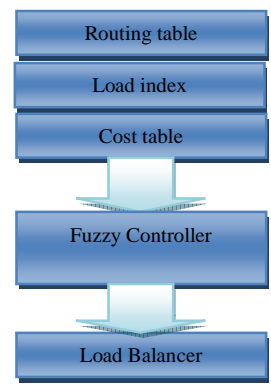

Fig.1:System model

The routing table presents the communication links among nodes in the system. Load index indicates the load of its related node, which is used by the policies in section II. In order to determine the node status as a sender, receiver or neutral by using fuzzy controller and based on fuzzy rules, we need a cost table that provides the nodes communication costs and the number of heavy loaded nodes. The cost table is obtained by using load index and routing table while the number of heavy loaded nodes can be extracted from the cost table.

### IV. METHODOLOGY

Load index value based on a given threshold is classified into five categories and is defined between 0 to w and threshold is s. Five Fuzzy sets (Fig.2) are used to describe the load index value: very lightly loaded, lightly loaded, moderate loaded, heavy loaded and very heavy loaded. Variables for load index take grade values of Fuzzy variables are uncertainties and depends on network situation it can be changed.

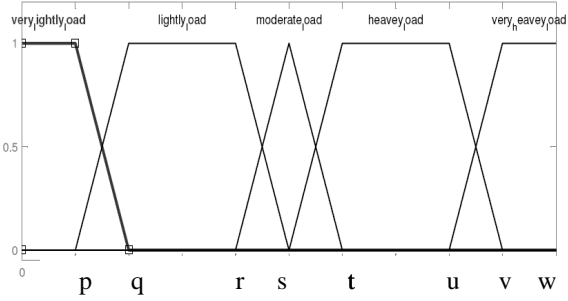

Fig.2: Fuzzy Index load chart

$$\mu_{verylightlyload}(\text{load}) = \begin{cases} 1 & \text{load} < p \\ \dfrac{q - \text{load}}{q - p} & p \leq \text{load} \leq q \\ 0 & \text{load} > q \end{cases}$$

$$\mu_{lightly\ load}(\text{load}) = \begin{cases} 0 & \text{load} < p\ \text{or} > s \\ \dfrac{\text{load} - p}{q - p} & p \leq \text{load} \leq q \\ 1 & q < \text{load} < r \\ \dfrac{s - \text{load}}{s - r} & r \leq \text{load} \leq s \end{cases}$$

$$\mu_{moderate\ load}(\text{load}) = \begin{cases} 0 & \text{load} < r\ \text{or} > t \\ \dfrac{\text{load} - r}{s - r} & r \leq \text{load} \leq s \\ \dfrac{t - \text{load}}{t - s} & s \leq \text{load} \leq t \end{cases}$$

$$\mu_{veryheavy\ load}(\text{load}) = \begin{cases} 0 & \text{load} < s\ \text{or} > v \\ \dfrac{\text{load} - s}{t - s} & s \leq \text{load} \leq t \\ 1 & t < \text{load} < u \\ \dfrac{v - \text{load}}{v - u} & u \leq \text{load} \leq v \end{cases}$$

$$\mu_{heavyload}(\text{load}) = \begin{cases} 1 & \text{load} < u \\ \dfrac{v - \text{load}}{v - u} & u \leq \text{load} \leq v \\ 0 & \text{load} > v \end{cases}$$

for input 2, number of heavy nodes fuzzy sets are define as less and more equal (N is number of heavy nodes).

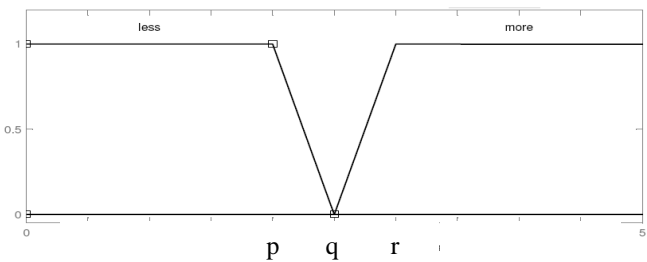

Fig.3: Fuzzy Input load

$$\mu_{less}(N) = \begin{cases} 1 & N < p \\ \dfrac{q - N}{q - p} & p \leq N \leq q \\ 0 & N > q \end{cases}$$

$$\mu_{moreequal}(N) = \begin{cases} 0 & N < q \\ \dfrac{N - q}{r - p} & q \leq N \leq r \\ 1 & N > r \end{cases}$$

Assuming sender initiated load balance algorithm, the proposed knowledge base is as follows:

*Rule [1]. If (load is very_lightly_load)*
*then (status__loadbalance__node is receiver)*

*Rule [2]. If (load is very_heavey_load)*
*then (status__loadbalance__node is sender)*

*Rule [3]. If (load is heavey_load) and*
*(no__heavy__load___nodes is more)*
*then (status__loadbalance__node is reciver)*

*Rule [4]. If (load is heavey_load) and*
*(no__heavy__load___nodes is less)*
*then (status__loadbalance__node is sender)*

*Rule [5]. If (load is lightly_load) and*
*(no__heavy__load___nodes is less)*
*then (status__loadbalance__node is sender)*

*Rule [6]. If (load is lightly_load) and*
*(no__heavy__load___nodes is more)*
*then (status__loadbalance__node is reciver)*

*Rule [7]. If (load is moderate_load) and*
*(no__heavy__load___nodes is more)*
*then (status__loadbalance__node is reciver)*

*Rule [8]. If (load is moderate_load) and*
*(no__heavy__load___nodes is less)*
*then (status__loadbalance__node is sender)*

*Rule [9] IF the node is sender*
*Then select a receiver as a migration partner*

*Rule [10] IF the node fails to find a migration partner*
*Then the node is neutral*

*Rule [11] IF the node is a sender*
*Then select a suitable task to transfer*

*Rule [12] IF the node fails to select a suitable task to transfer*
*Then select another migration partner*

Fuzzy sets for output are shown as Fig.4:

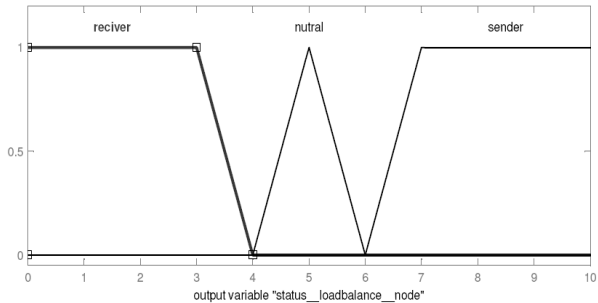

Fig.4: Fuzzy output

## V. PERFORMANCE EVALUATION

Simulation was performed in MATLAB and NS2 to verify our approach. We evaluate our fuzzy load balancer in a system with five node using a randomly generated network graph and a random generated load vector-load vector consist of the number of task on the node and load index for each node. The edge connectivity in the network graph is generated with probability of 0.2 and task allocation with a Uniform distribution U [0, 1]. The generated task is assigned to the node corresponding to the interval of the generated random variable. Inter arrival times are taken from the exponential distribution. Processor speeds for all nodes are taken from Uniform distribution. Our fuzzy proposed algorithm in form of real time during updating amount of nodes load refreshes the cost table. Then we generated the cost table according to network graph and load vector. Load of each node is equal to the number of the node tasks. From cost table we can calculate the number of heavy nodes. In fuzzy system according to status of heavy load nodes, amount of node load and based on fuzzy rule base, we can determine status of each node, which can be in one of three states: sender, receiver and neutral. Results of our fuzzy load balancer algorithm are presented in Table 1.

Table 1: Response time of load balancing algorithm for different number of tasks.

| Algorithm | Number of Task | | | | |
|---|---|---|---|---|---|
| | 2 | 4 | 6 | 8 | 10 |
| Randomize | 3 | 4 | 7 | 11 | 16 |
| Round Robin | 2 | 3 | 6 | 9 | 13 |
| Fuzzy | 1 | 2 | 4 | 7 | 11 |

Fig. 5 shows fuzzy approach has significantly better response time. In Table 2 improvement percentage of our algorithm for different number of tasks in Round Robin and Randomize algorithm are shown. This table shows performance of fuzzy algorithm is better than RR and randomize algorithm.

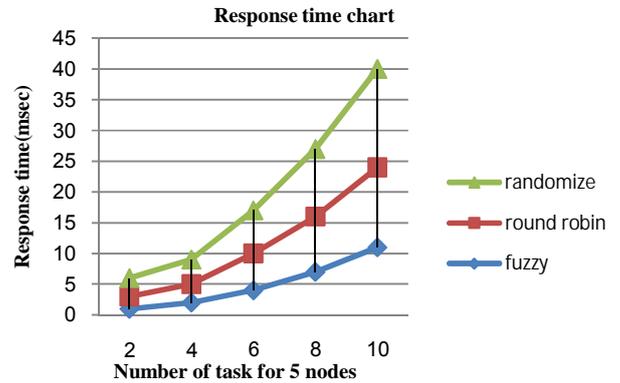

Fig.5: Response Time between Randomize, Round Robin and Fuzzy load balancing algorithm.

Table 2: Comparison of improvement Percentage of Fuzzy purpose algorithm vs. Round Robin & Randomize load-balancing algorithm.

| Number Of Task | Performance Of Fuzzy vs. Round Robin | Performance Of Fuzzy vs. Randomize |
|---|---|---|
| 2 | %50 | %66.7 |
| 4 | %33.3 | %50 |
| 6 | %33.3 | %42.9 |
| 8 | %22.2 | %36.4 |
| 10 | %15.4 | %31.25 |

In Table 3 total improvement of our fuzzy approach is shown. This table confirms fuzzy load balancing algorithm has better response time and performance in comparison to Round Robin and Randomize load balancing algorithm respectively 30.84% and 45.45%.

Table 3: proportion percentage of improving our novel algorithm

| | Round Robin | Randomize |
|---|---|---|
| Fuzzy | % 30.84 | % 45.45 |

## CONCLUSION AND FUTURE WORKS

Fuzzy logic systems can make absolute outputs from uncertainties inputs. In this paper, we present a new approach for implementing dynamic load balancing algorithm with fuzzy logic and we have shown its response time is significantly better than round robin and randomize algorithm.

In the future works, we will follow the load balancing issue in parallel systems to find out whether the load balancing action will be quicker than the previous works or not. Moreover, we will present a new load balancing approach for predicting the nodes status as sender, receiver or neutral with less time complexity by using genetic algorithms and neurofuzzy techniques.

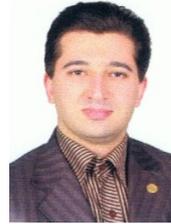

**Abbas Karimi**: Received his Bachelor degree in Computer hardware engineering and MS in Computer Software Engineering from Iran. He is PhD candidate in UPM, Malaysia in the field of computer system Engineering. He has been working as a lecturer and faculty member in the Department of computer engineering at IAU-Arak Branch and lecturer in several universities. He was involved in several research projects, authorizing one textbook in Persian, several management posts, etc. His research interests are in load balancing algorithms, real time, distributed, parallel and fault-tolerant systems.